\documentclass{ieeetj}
\usepackage{cite}
\usepackage{amsmath,amssymb,amsfonts}
\usepackage{algorithmic}
\usepackage{graphicx,color}
\usepackage{textcomp}
\usepackage[dvipsnames]{xcolor}
\usepackage{hyperref}
\hypersetup{hidelinks=true}
\usepackage{algorithm,algorithmic}
\def\BibTeX{{\rm B\kern-.05em{\sc i\kern-.025em b}\kern-.08em
    T\kern-.1667em\lower.7ex\hbox{E}\kern-.125emX}}
\AtBeginDocument{\definecolor{tmlcncolor}{cmyk}{0.93,0.59,0.15,0.02}\definecolor{NavyBlue}{RGB}{0,86,125}}
\usepackage{makecell}

\usepackage{booktabs}
\usepackage{multirow}
\usepackage{subcaption}
\usepackage{pgfplots}
\usetikzlibrary{pgfplots.groupplots, decorations.pathreplacing, calc}

\def\OJlogo{\vspace{-4pt}$<$Society logo(s) and publication title will appear here.$>$}
\def\seclogo{\vspace{10pt}$<$Society logo(s) and publication title will appear here.$>$}

\def\authorrefmark#1{\ensuremath{^{\textbf{#1}}}}

\usepackage{titlesec}
\usepackage{lipsum}

\titlespacing*{\subsubsection}{-0.2pt}{0.5ex plus 0.5ex minus .1ex}{0.5ex plus .5ex}

\makeatletter
\renewcommand*{\p@subsection}{\thesection.}
\makeatother

\begin{document}
\receiveddate{XX Month, XXXX}
\reviseddate{XX Month, XXXX}
\accepteddate{XX Month, XXXX}
\publisheddate{XX Month, XXXX}
\currentdate{XX Month, XXXX}
\doiinfo{XXXX.2022.1234567}

\markboth{}{Author {et al.}}
\title{VoicePAT: An Efficient Open-source Evaluation Toolkit for Voice Privacy Research}

\author{Sarina Meyer\authorrefmark{1}*, Student Member, IEEE, and Xiaoxiao Miao\authorrefmark{2,3}*, Member, IEEE\\  and Ngoc Thang Vu\authorrefmark{1}, Member, IEEE}
\affil{Institute for Natural Language Processing, University of Stuttgart, Germany}
\affil{Singapore Institute of Technology, Singapore}
\affil{National Institute of Informatics, Japan}
\corresp{Corresponding author: Sarina Meyer (email: sarina.meyer@ims.uni-stuttgart.de).}
\authornote{* Equal contribution. This study is supported by the Carl Zeiss foundation, JST CREST Grants (JPMJCR18A6), and MEXT KAKENHI Grant (22K21319).}

\begin{abstract}
Speaker anonymization is the task of modifying a speech recording such that the original speaker cannot be identified anymore. Since the first Voice Privacy Challenge in 2020, along with the release of a framework, the popularity of this research topic is continually increasing. However, the comparison and combination of different anonymization approaches remains challenging due to the complexity of evaluation and the absence of user-friendly research frameworks.
We therefore propose an efficient speaker anonymization and evaluation framework based on a modular and easily extendable structure,
almost fully in Python. The framework facilitates the orchestration of several anonymization approaches in parallel and allows for interfacing between different techniques. Furthermore, we propose modifications to common evaluation methods which improves the quality of the evaluation and reduces their computation time by 65 to 95\%, depending on the metric. Our code is fully open source.
\end{abstract}  


\maketitle

\begin{figure*}
    \centering
    \includegraphics[width=\textwidth]{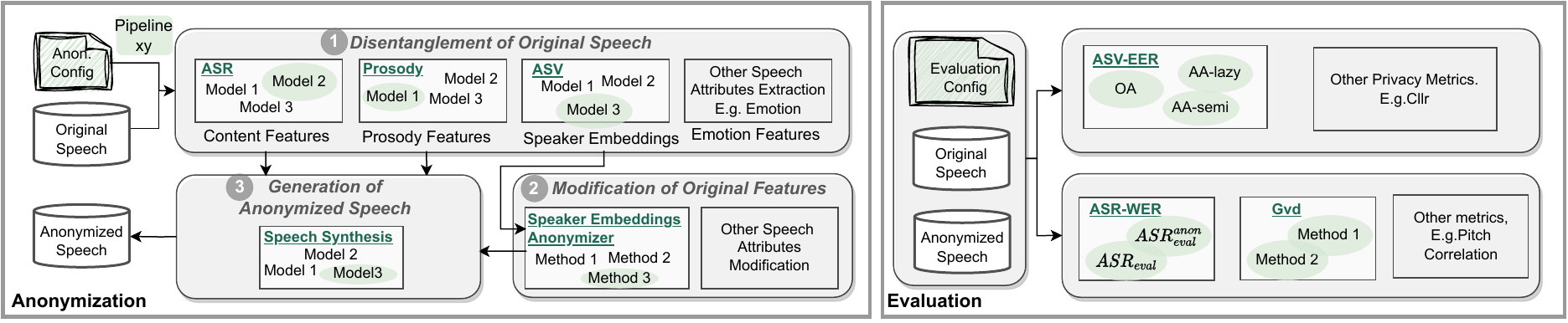}
    \caption{Proposed framework, consisting of separate anonymization and evaluation branches and example configurations for each branch}.
    \label{fig:framework}
\end{figure*}

\section{INTRODUCTION} 
\IEEEPARstart{S}{peaker} anonymization \cite{tomashenko2021voiceprivacy} is a task in which speech recordings are automatically modified such that the original speaker becomes unidentifiable from the audio, usually by changing the voice in the direction of an artificial target speaker. The goal of this process is to preserve the speaker's voice privacy while keeping enough information from the input to use the anonymized audio in downstream tasks (e.g., speech recognition \cite{povey2018semi}).
In order to foster research in this topic, the Voice Privacy Challenge (VPC) has been held in 2020 and 2022 \cite{tomashenko2021voiceprivacy,tomashenko2022voiceprivacy}. The open-source framework accompanying the challenges \textemdash consisting of code bases, baseline and evaluation models, datasets, and techniques \textemdash have had a great influence on the field, with most approaches using at least part of the VPC framework.

The primary goal of the VPC framework is to address the challenges associated with voice privacy research.
However, it has two key drawbacks: (1) The heavy reliance on the Kaldi toolkit \cite{kaldi} results in processes that are complicated and opaque. (2) The static structure of this framework introduces many redundant computations, leading to inefficiencies.

Given the current importance of voice privacy research in society and politics, there is demand for a framework that allows quick ideation and experimentation without the burden of infrastructure issues, in order to promote collaboration among researchers. Although alternative evaluation frameworks \cite{franzreb2023comprehensive, zhang2023voicepm} exist, they do not adhere to the standard evaluation protocol introduced by the VPC. 
We thus propose a robust and modular framework for speaker anonymization. The framework is implemented almost fully in Python and consists of two branches (1) for anonymization and (2) evaluation, as shown in Figure \ref{fig:framework}. Both branches exhibit a modular structure in which single components can be skipped or added. The methods and models in each component can be easily exchanged for alternatives. The control over the composition of a pipeline is done exclusively via configuration files such that different speaker anonymization systems (SASs) or evaluation metrics can be compared with minimal effort. 
We extend and improve commonly used evaluation methods by employing the ESPnet \cite{espnet} and SpeechBrain \cite{speechbrain} toolkits for evaluation instead of Kaldi \cite{kaldi}. Furthermore, we significantly speed up the evaluation by combining data reduction and finetuning techniques for privacy and utility metrics. 
Overall, we believe this framework will be an essential tool for advancing research in voice privacy and facilitating participation in future Voice Privacy Challenges in the field. 

Our contributions are as follows:
\begin{itemize}
    \item We propose improvements to standard evaluation methods for speaker anonymization which make these evaluations more efficient, easier to use, and more feasible for intermediate evaluations.
    \item We show through experiments on 
    two primary VPC baselines and two state-of-the-art SASs that these improvements provide stronger privacy tests, while requiring 65 to 95\% less time to execute. 
    \item We release all code in a new toolkit, 
    VoicePAT\footnote{\url{https://github.com/DigitalPhonetics/VoicePAT}} (Voice Privacy Anonymization Toolkit), which further includes pipelines for running different state-of-the-art SASs. 
    \item Due to its modularity, extending VoicePAT with more anonymization systems and evaluation metrics is easily possible, enabling the comparison of different approaches within one framework and probing the effectiveness of the anonymization using a more diverse attack landscape.
\end{itemize}
 
\section {Background and Related Work}
Before describing our proposed framework and evaluation, we will first give some background about the current state of the topic, existing evaluation metrics, and frameworks.

\subsection{SPEAKER ANONYMIZATION} \label{subsec:anonymization}
The goal of an SAS \cite{tomashenko2021voiceprivacy, tomashenko2022voiceprivacy} is to automatically modify the voice in an audio recording to make the original speaker unidentifiable. 
Following VPC baseline models, approaches can be categorized into methods based on disentanglement and those based on digital signal processing.
Whereas the latter perceptively modify speech to conceal the original speaker's identity \cite{tomashenko2021voiceprivacy,tomashenko2022voiceprivacy}, they are less effective than disentanglement-based methods \cite{meyer22b_interspeech, meyer2023anonymizing, meyer2023prosody, miao22_odyssey,miao2022analyzing,miao2023language,mawalim2022speaker,champion2022disentangled, shamsabadi2022differentially,turner2022generating,yao22_spsc}, which involve three steps as shown in the left of Figure \ref{fig:framework}:

\noindent
\textit{(1) Disentanglement of Original Speech}: 
(i) Frame-level content features extraction via automatic speech recognition (ASR) models \cite{povey2018semi, peddinti2015time} or self-supervised learning (SSL) based content encoders \cite{miao22_odyssey, miao2022analyzing};
(ii) Frame-level prosody features extraction, e.g., F0 using the YAAPT algorithm \cite{kasi2002yet};
(iii) Utterance-level speaker embeddings extraction, e.g., from pre-trained x-vector \cite{snyder2018x} or ECAPA-TDNN \cite{desplanques2020ecapa} models for automatic speaker verification (ASV).

\noindent
\textit{(2) Modification of original features}: 
This is a crucial step to hide the original speaker's identity. 
Most works focus on modifying the original speaker embeddings, assuming that identity is mainly encoded in them.
Typically, a selection-based speaker anonymizer \cite{Srivastava2020DesignCF}, 
replaces an original speaker with an anonymized speaker vector.
This anonymized vector is a mean speaker vector derived from a randomly chosen set of speaker vectors in an external pool.
However, the averaging process on speaker vectors leads to limited speaker diversity in the generated anonymized voices and serious speaker privacy leakage problems when facing stronger attacker \cite{miao2023language}. 
To mitigate these problems, recent works adopt DNN-based anonymizers.
For example, \cite{meyer2023anonymizing, meyer2023prosody} introduce a Wasserstein generative adversarial network (\textbf{GAN}) that is trained to turn random noise into artificial speaker embeddings that follow a similar distribution as the original speaker vectors.
Another approach employs an orthogonal householder neural network (\textbf{OHNN}) to rotate original speaker vectors, ensuring that anonymized vectors follow the original space and maintaining speech naturalness. The parameters of the OHNN-based anonymizer are trained using classification and similarity losses, encouraging distinct speaker identities \cite{miao2023language}.

\noindent
\textit{(3) Generation of anonymized speech}: 
The anonymized speaker, prosody, and content features are then fed into a speech synthesis model \cite{wang2019neural,kong2020hifi} to generate anonymized speech.

The VPC introduced two primary disentanglement-based SASs. In \textbf{BL 1.a}, speech is disentangled into speaker identity by a pre-trained x-vector, fundamental frequency by YAAPT algorithm, and linguistic information by a pre-trained factorized time delay neural network (TDNN-F) based ASR acoustic model \cite{povey2018semi, peddinti2015time}. Then, the selection-based speaker anonymization scheme \cite{Srivastava2020DesignCF} modifies a source x-vector to hide speaker information. The speech synthesis acoustic model (SS AM) generates Mel-filterbank features using the anonymized pseudo x-vector, F0, and linguistic features, followed by a neural source-filter (NSF)-based waveform generation model \cite{wang2019neural} to synthesize anonymized speech.
Similar to \textbf{BL 1.a}, \textbf{BL 1.b} replaces the traditional speech synthesis pipeline (SS AM + NSF) with a unified HiFi-GAN \cite{kong2020hifi} NSF model as the waveform generator.

\subsection{EVALUATION} {\label{sec:related}}
Speaker anonymization commonly has two objectives: (1) \textit{privacy:} protecting the identity of a speaker, and (2) \textit{utility:} keeping other attributes of the original audio needed for use in downstream applications (e.g., linguistic content, naturalness, prosody, speaker emotion) unchanged. 
The challenge is to optimize an SAS to achieve a trade-off between both objectives,
whereby the weighting between them and the utility assessment metrics depend on the application.

\subsubsection{SPEAKER PRIVACY PROTECTION}
To evaluate the effectiveness of preserving speaker identity against different attackers, it is most common to compute the equal error rate (EER) using ASV evaluation models. For this, an attacker compares the anonymized trial utterance processed by users against enrollment utterances in different attack conditions with either a model trained on original ($ASV_\text{eval}$) or on anonymized ($ASV_\text{eval}^\text{anon}$) data \cite{tomashenko2021voiceprivacy, tomashenko2022voiceprivacy}:
\begin{itemize}
\item \textit{Unprotected} (OO): a baseline metric to assess the effectiveness of the $ASV_\text{eval}$ attacker when both enrollment and trial utterances are not anonymized (i.e., original).
\item \textit{Ignorant} (OA): attackers are unaware of the anonymization, they use original enrollment data and $ASV_\text{eval}$ to infer the identity of anonymized trial utterances.
\item \textit{Lazy-informed} (AA-lazy): attackers use anonymized enrollment speech, generated with the same SAS but inaccurate parameters, and $ASV_\text{eval}$ to detect identities.
\item \textit{Semi-informed} (AA-semi): the attacker is similar to the \textit{lazy-informed} one, but employs a more powerful $ASV_\text{eval}^\text{anon}$ model trained on anonymized speech, which helps to reduce the mismatch between the original and anonymized speech to infer the speaker identity.
\end{itemize} 
For a successful anonymization, an EER close to 50\% is targeted.
An alternative to the EER metric is using the log-likelihood-ratio cost function $C_{llr}$ as discrimination loss ($C_{llr}^{min}$) and calibration loss ($C_{llr} - C_{llr}^{min}$) \cite{tomashenko2021voiceprivacy}. Other privacy metrics include the linkability ($D_{\leftrightarrow}^{sys}$) between two utterances \cite{gomez-barrero2018general, maouche2020a-comparative}, the de-identification ($De_{ID}$) based on voice similarity matrices \cite{noe2020speech}, and the expected ($D_{ECE}$) and worst-case ($\log(l)$) privacy disclosure metrics of the ZEBRA framework \cite{nautsch2020zebra}. 

\subsubsection{SPEECH UTILITY PRESERVATION}
As most applications require the anonymization to retain the linguistic content of the speech, the primary evaluation for speech utility is performed with ASR and measured as word error rate (WER). The VPC introduces two models for this: $ASR_\text{eval}$ (A-lazy) is trained on the original data, and $ASR_\text{eval}^\text{anon}$ (A-semi) is trained on the anonymized data.  
$ASR_\text{eval}^\text{anon}$ tests the best-case scenario in which the downstream ASR knows how anonymization affects the audio quality and can adapt to it, whereas $ASR_\text{eval}$ might simulate a more realistic condition in which such information is not known.
The lower the WER is, the better.

Another common utility metric is the gain of voice distinctiveness $\text G_\text{VD}$ \cite{maouche2020a-comparative} that assesses how well the ability of distinguishing different speakers is kept during anonymization. If the voice distinctiveness in the anonymized space is the same as in the original space, the $\text G_\text{VD}$ is close to zero. If it is improved, the score is above zero, otherwise below. The metric is closely related to the privacy metrics and is computed using $ASV_\text{eval}$ and voice similarity matrices.

Depending on the application, other speech utility metrics could be included, e.g., prosody preservation via pitch correlation as done in the VPC 2022 \cite{tomashenko2022voiceprivacy}. A simple approach could be to transcribe speech with an ASR model and synthesize it back to speech from the transcription using a text-to-speech (TTS) system. This conceals speaker identity but also removes other paralinguistic features like emotion and health status which are important for health applications.

For simplicity, we will focus on EER as a privacy indicator, and WER and $\text G_\text{VD}$ for measuring utility in speech recognition tasks in our experiments.

\subsection{EXISTING FRAMEWORKS AND THEIR LIMITATIONS}
To support participation in the challenges, the VPC published an open-source framework with code for all baselines and evaluation metrics. However, this framework lacks the flexibility to skip single steps in its run pipeline, rerun only parts of it, or add new metrics. Most of the algorithms are written in the C++-based Kaldi toolkit which is challenging to maintain and lacks compatibility with standard Python-based speech processing models. Furthermore, the evaluation models included in the framework can take several days for computations. 
Combined with the difficulty of skipping previously computed calculations, performing a full anonymization with subsequent evaluation in the VPC framework is complicated and expensive, potentially discouraging new researchers from working on SAS development. 
Motivated by similar concerns about the framework, \cite{franzreb2023comprehensive,zhang2023voicepm} recently presented an alternative evaluation framework written in Python and exhibiting modular and extendable structures. However, they do not test their framework with standard SAS approaches nor compare their evaluation metrics with the ones in the VPC framework. This makes it difficult to assess the quality of their improvements. 
We thus find a lack of suitable anonymization and evaluation frameworks for this topic. Hence, we propose a new alternative in this paper.

\section{PROPOSED EFFICIENT FRAMEWORK}

\begin{figure*}[!t]
\centering
\subfloat[][Privacy]{\includegraphics[height=2.6cm]{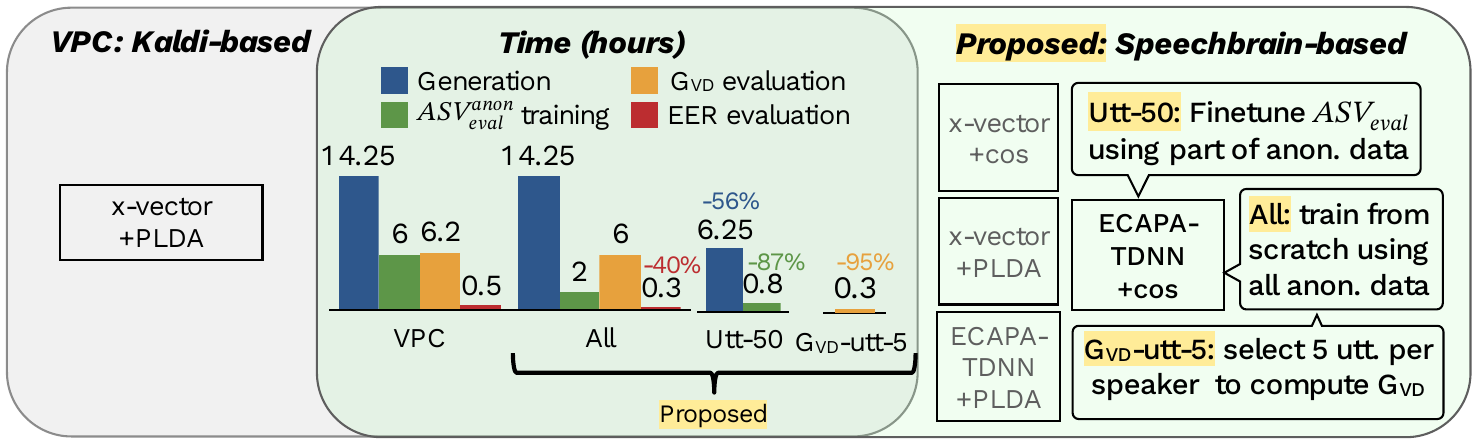}\label{fig:compare-vpc-proposed-privacy}}
\hspace{0.2cm}
\subfloat[][Utility]{\includegraphics[height=2.6cm]{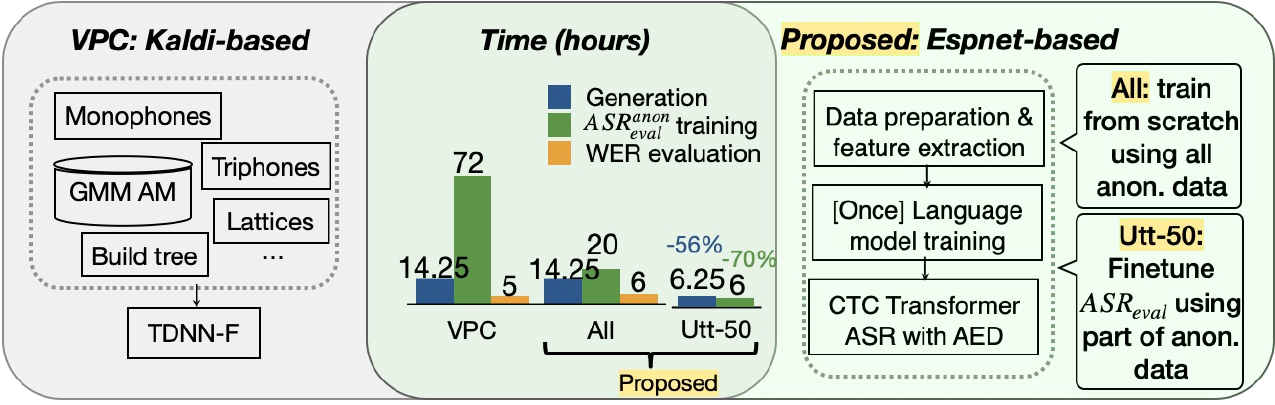}\label{fig:compare-vpc-proposed-utility}}\vspace{0.0cm}\\
\caption{Comparisons of VPC and proposed privacy and utility evaluation models. Note that although $\text G_\text{VD}$ is a utility metric, we plotted $\text G_\text{VD}$ evaluation time in the privacy subplot as it is computed by $ASV_{eval}$.}
\label{fig:compare-vpc-proposed}
\end{figure*}

The proposed framework for speaker anonymization research consists of two pipeline branches, shown in Figure \ref{fig:framework}: One for the anonymization process and one for evaluation. Both branches consist of several modules, which can be instantiated with different models. 
All parameters for selecting the order and type of modules, 
as well as all other settings for running a pipeline, are given in configuration files. Modules can thus be exchanged, extended, or skipped if the general objective allows it. In this way, the framework provides a flexible option to modify existing approaches, combine ideas from different systems, and test the effect of single components in a controlled fashion.

\subsection{ANONYMIZATION BRANCH}
The goal of the anonymization branch is to provide a platform for researchers developing an SAS to evaluate their ideas quickly. Ideally, if they only want to test a minor change like a different speaker embedding modification mechanism, they would only have to add a new model (a Python class) to the speaker embedding modification module and adjust the configuration file. If the modification is radical, they might need to add a new module and pipeline. 

Generally, an SAS consists of the following components: (1) a configuration file, (2) a pipeline, (3) a collection of modules, and (4) one specific model in each module. The configuration file specifies the pipeline (e.g., the GAN-based pipeline of \cite{meyer2023prosody}), which then defines the obligatory and optional modules and their processing order. 
Each module can be instantiated with different models or approaches, e.g., different speech synthesis models. 
The selection of one model per module and the inclusion of optional modules are specified in the configuration file. 
Per default, the output of each module is saved to disk. This makes it possible to skip the computation of one module if it has been computed before and its input has not changed, and thus, to test minor modifications more efficiently.

\subsection{EVALUATION BRANCH}
Following the standards of voice privacy research, the evaluation branch is divided into two modules: privacy and utility. Each module corresponds to one evaluation aspect and can consist of one or several metrics. For example, ASV is one module of privacy evaluation and is mainly measured by one metric, EER, however, multiple models can be used to calculate this. Similar to the anonymization branch, all settings are again set in configuration files. 
To further improve the efficiency of the proposed framework, we employ more powerful ASV and ASR models, explore the training strategies, and modify the computation of $\text G_\text{VD}$ as described below.

\subsubsection{Evaluation models}
In the VPC framework, EER and $\text G_\text{VD}$ are computed using a Kaldi-based x-vector speaker encoder with a PLDA distance model \cite{snyder2018x}, and WERs are computed using a Kaldi-based TDNN-F model \cite{povey2018semi}. 
As ASV and ASR technology develops, it is important to examine the impact of advanced models on the respective evaluation results. Aiming to find the most reliable choice, we propose using evaluation models based on the state-of-the-art toolkits Speechbrain \cite{speechbrain} for ASV and ESPnet \cite{espnet} for end-to-end ASR, as shown in Figure \ref{fig:compare-vpc-proposed}.

Both toolkits are developed using PyTorch \cite{pytorch} in the context of research, and they meet two requirements for our purposes: (1) They provide a user-friendly approach to modify training recipes which cover a wide range of hyperparameters and architecture choices for the models. (2) Both toolkits are continuously developed, ensuring they incorporate the latest advancements in both ASV and end-to-end ASR techniques regularly.
For ASV evaluation models, we present choices including the Speechbrain-based  x-vector and the cutting-edge ECAPA-TDNN, featuring both cosine and PLDA back-ends.
For ASR evaluation models, we provide an ESPnet-based Transformer encoder-based Connectionist Temporal Classification (CTC) ASR model  with Attention Encoder Decoder (AED) \cite{watanabe2017hybrid}.
A transformer-based language model is trained using \textit{LibriSpeech-train-clean-360} \cite{panayotov2015librispeech} once and used for decoding.

\subsubsection{Modifications for $\text{G}_\text{VD}$ }
The gain of voice distinctiveness metric $\text G_{\text{VD}}$ \cite{noe2020speech} is defined as the diagonal dominance ratio of two voice similarity matrices, one for the original speaker space and one for the anonymized one. 
In the VPC framework, those similarity scores are computed by the $ASV_\text{eval}$ model trained on the original data.
$ASV_\text{eval}$ yields more reliable scores for the original data but introduces a mismatch when applied to anonymized speech. 
We are interested in exploring different evaluation models for computing similarity scores:
(1)  using the $ASV_\text{eval}$ model;
(2)  using the $ASV_\text{eval}^{anon}$ model;
(3)  using $ASV_\text{eval}$ for original data and $ASV_\text{eval}^{anon}$ for anonymized data to see whether this could improve the accuracy of similarity scores for different types of data. Furthermore, different from the VPC framework, where all utterances of each speaker are considered for similarity computation, we enhance efficiency by randomly selecting 5 utterances per speaker to compute the log-likelihood ratios between two speakers.

\subsubsection{Training strategy of $ASV_\text{eval}^\text{anon}$ and $ASR_\text{eval}^\text{anon}$ models}
\label{sec:training strategy}
Four models are required for privacy and utility evaluation, as described in Section \ref{sec:related}: $ASV_\text{eval}$ and $ASR_\text{eval}$, trained on the original \textit{LibriSpeech-train-clean-360} dataset, are directly provided by the VPC platform. Thus, we will assess only their evaluation times, not the training time.
$ASV_\text{eval}^\text{anon}$ and $ASR_\text{eval}^\text{anon}$ are trained from scratch like the original models,
but on the anonymized \textit{LibriSpeech-train-clean-360} processed by the same evaluated SAS.
This requires extra time for anonymizing training data and conducting the training process, in addition to the evaluation time.

Since the anonymization of the entire training dataset is quite time-consuming, we aim to explore the impact of obtaining $ASV_\text{eval}^\text{anon}$ and $ASR_\text{eval}^\text{anon}$ by finetuning the pretrained $ASV_\text{eval}$ and $ASR_\text{eval}$, respectively, using only a subset of the anonymized data 
to eliminate the necessity of anonymizing the entire dataset.
We consider two techniques for data reduction: (1) choosing a limited number of utterances from all the speakers, and (2) selecting all utterances from a specific subset of speakers. 
By using this approach, we can balance the trade-off between anonymization, training time, and the effectiveness of the $ASV_\text{eval}^\text{anon}$ and $ASR_\text{eval}^\text{anon}$.

\section{EXPERIMENTS} 
In this section, we compare the effectiveness of different evaluation models in measuring privacy and utility performance across various SASs. The evaluation models used here are trained on either the entire original or anonymized \textit{LibriSpeech-train-clean-360} dataset.
Once the evaluation models are chosen, the focus shifts to exploring the training strategy for the \textit{semi-informed} evaluation models, particularly concerning the amount of training data. 

To draw more general conclusions, we choose diverse disentanglement-based SASs, including both traditional and state-of-the-art methods, to generate anonymized speech and evaluate it using our proposed VoicePAT. The specifications for these SASs are listed in Table \ref{tab:notations}. All SASs follow the three steps described in Section \ref{subsec:anonymization} with different realizations of each component and different anonymization techniques on the speaker embeddings.
All experiments are performed on the VCTK \cite{yamagishi2019vctk} and LibriSpeech \cite{panayotov2015librispeech} test sets as given by the VPC. The results consistently report the average score on them. Further settings, e.g., training hyperparameters, can be found in our source code. All time measurements apply to experiments conducted on single NVIDIA A100 GPU, except ASR evaluation using 4 GPUs.

\begin{table}[t]
\centering
\caption{Specifications for the evaluated SASs. GST means global style token \cite{meyer2023prosody}, ROH means random orthogonal Householder \cite{miao2023language}.}
\label{tab:notations}
\resizebox{1\linewidth}{!}{
\begin{tabular}{lcccc}
\toprule
  System & \makecell[c] {Content \\ encoder} &  \makecell[c]{Speaker  \\encoder}  & \makecell[c]{Speech\\ synthesis}   & \makecell[c] {Speaker \\ anon.} \\\midrule
  {\textbf{BL 1.a} \cite{tomashenko2022voiceprivacy}}               & {TDNN-F}  & {x-vector}   &{SS-AM+NSF}  & {Select.}   \\  
    {\textbf{BL 1.b} \cite{tomashenko2022voiceprivacy}}               & {TDNN-F}  & {x-vector}   &{HiFi-GAN+NSF}  & {Select.}   \\ 
    {\textbf{GAN} \cite{meyer2023prosody}}               & {Branchformer+CTC/AED }  & {GST-based}   & {FastSpeech2 + HiFi-GAN}  & {GAN}   \\ 
   \textbf{OHNN} \cite{miao2023language}             & {SSL}  & {ECAPA}   & {HiFi-GAN}  &   \makecell[c]{ROH}    \\     
\bottomrule
\end{tabular}
}
\end{table}

\subsection{CHOICE OF EVALUATION MODELS}
\subsubsection{ASV evaluation models}
\begin{table}[]
    \centering
    \caption{Comparison of four privacy attack models using the x-vector and ECAPA speaker encoders with PLDA and cosine as distance measures. Privacy scores for each SAS and attack condition are given as EER in \%. $\uparrow$ means higher values are better, while $\downarrow$ means lower values are better.}
    \resizebox{\linewidth}{!}{
    \begin{tabular}{cc|rrrr}
        \toprule
        SAS & Eval  & OO $\downarrow$ & OA $\uparrow$ & AA-lazy $\uparrow$ & AA-semi $\uparrow$\\
        \midrule
        \multirow{4}{*}{\textbf{BL 1.a}} & x-vector + cosine & 9.24 & 54.38 & 26.39 & 11.73\\
        & x-vector + PLDA & 7.18 & 52.56 & 26.41 & 8.66\\
        & ECAPA + cosine & 3.11 & 52.85 & 25.65 & 7.62\\
        & ECAPA + PLDA & 3.49 & 51.62 & 25.06 & 7.69\\
        \midrule
        \multirow{4}{*}{\textbf{BL 1.b}} & x-vector + cosine &  9.24 & 52.30 & 25.31 & 11.33\\
        & x-vector + PLDA & 7.18 & 51.49 & 25.98 & 8.27\\
        & ECAPA + cosine & 3.11 & 49.82 & 24.67 & 7.21\\
        & ECAPA + PLDA & 3.49 & 50.84 & 23.10 & 8.51\\
        \midrule
        \multirow{4}{*}{\textbf{OHNN}} & x-vector + cosine & 9.24 & 49.65 & 45.55 & 47.09\\
        & x-vector + PLDA & 7.14 & 49.34 & 45.55 & 42.47\\
        & ECAPA + cosine & 3.11 & 48.22 & 44.26 & 44.62\\
        & ECAPA + PLDA & 3.63 & 50.18 & 44.44 & 41.47\\
        \midrule
        \multirow{4}{*}{\textbf{GAN}} & x-vector + cosine & 9.24 & 53.05 & 48.57 & 43.11\\
        & x-vector + PLDA & 7.03 & 52.80 & 48.71 & 44.12\\
        & ECAPA + cosine & 3.11 & 51.45 & 46.11 & 44.61\\
        & ECAPA + PLDA & 3.68 & 50.85 & 47.57 & 47.20\\
        \bottomrule
    \end{tabular}}
    \label{tab:ecapa_x-vector}
\end{table}

Table \ref{tab:ecapa_x-vector} lists the mean EERs for various SASs under all conditions computed by different evaluation models. 
First, no matter which ASV attacker is used, the EERs of \textbf{BL 1.a} and \textbf{BL 1.b} decrease by around 25\% under the AA-lazy and 7\%-11\% under the AA-semi condition, indicating severe privacy leakage when facing the stronger, \textit{semi-informed} attack model $ASV_\text{eval}^\text{anon}$.
However, the EERs of \textbf{OHNN}- and \textbf{GAN}-based SASs yield over 40\% across all the attack conditions, showing remarkable privacy protection capabilities.
In order to decide which evaluation model provides the best results, we look at the OO condition, where no SAS is applied and it is expected to achieve very low EERs on this original data.
The model using ECAPA-TDNN and cosine distance achieves with 3.11\% the lowest EER and can therefore be considered as the best ASV evaluation model.
This is consistent with the findings in the ASV field \cite{wang22r_interspeech,li2023make}.

\begin{table}[]
    \centering
    \caption{EERs and WERs obtained by the proposed and the VPC evaluation models. The proposed models are ECAPA-TDNN + cosine for ASV and transformer-based CTC/AED ASR. \textit{-l} and \textit{-s} stand for \textit{-lazy} and \textit{-semi}.}
        \resizebox{\linewidth}{!}{
    \begin{tabular}{cc|rrrr|rrr}
        \toprule
        & & \multicolumn{4}{c|}{EER (\%) $\uparrow$} & \multicolumn{3}{c}{WER (\%) $\downarrow$}\\
        SAS & Eval & OO $\downarrow$ & OA & AA-l & AA-s & O & A-l & A-s\\
        \midrule
        \multirow{2}{*}{\textbf{BL 1.a}} & VPC & 3.58 & 50.07 & 30.31 & 10.31 & 8.48 & 8.96 & 7.79\\ 
        & Proposed & 3.11 & 52.85 & 25.65 & 7.62 & 7.24 & 8.00 & 7.66\\ 
        \midrule
        \multirow{2}{*}{\textbf{BL 1.b}} & VPC & 3.58 & 51.11 & 28.98 & 9.93 & 8.48 & 10.16 & 7.61\\ 
        & Proposed & 3.11 & 49.82 & 24.67 & 7.21 & 7.24 & 8.42 & 7.50\\ 
        \midrule
        \multirow{2}{*}{\textbf{OHNN}} & VPC & 3.58 & 49.93 & 45.94 & 42.15 & 8.48 & 10.24 & 8.01 \\ 
        & Proposed & 3.11 & 48.22 & 44.26 & 48.61 & 7.24 & 7.92 & 7.70 \\ 
        \midrule
        \multirow{2}{*}{\textbf{GAN}} & VPC & 3.58 & 51.09 & 46.49 & 44.33 & 8.48 & 8.47 & 6.86\\ 
        & Proposed  & 3.11 & 51.45 & 46.11 & 48.30 & 7.24 & 8.00 & 7.03\\ 
        \bottomrule
    \end{tabular}}
    \label{tab:res_eval_models_summary}
\end{table}
 
 Accordingly, we choose ECAPA-TDNN + cosine as the primary proposed ASV evaluation model in all following experiments. In the central columns in Table \ref{tab:res_eval_models_summary}, we compare the EERs with this proposed model to the x-vector + PLDA model of the VPC toolkit.
 Compared to the VPC model, the proposed one consistently achieves lower or similar EERs across all the conditions and SASs. 
 This means that the proposed ASV model is a stronger attacker,
 which is reasonable as the proposed model exhibits a more powerful ability to infer the speaker's identity.

Moreover, the proposed ASV model reduces the time needed for it to train and perform the evaluation (Figure \ref{fig:compare-vpc-proposed-privacy}). Instead of 6 hours for training the VPC model (which includes the x-vector encoder and the PLDA), our proposed ECAPA-TDNN + cosine model only requires 2 hours. The effect on evaluation time is smaller, though still noticeable: reducing the time needed from 30 minutes to 20.
 
 \subsubsection{ASR evaluation models}
The right columns of Table \ref{tab:res_eval_models_summary} summarize the WERs for both original (O) and anonymized data (A-lazy, A-semi),
using the VPC (TDNN-F) or proposed (Transformer-based CTC/AED) evaluation models.
One common trend for all SASs is that the proposed ASR model achieves notably lower WERs in comparison to the VPC model. 
For the A-semi condition, the utilization of $ASR_\text{eval}^\text{anon}$ can reduce the mismatch between original and anonymized data, further decreasing the WERs.

Another interesting observation is that for the VPC model, the WERs of A-semi decoded by $ASR_\text{eval}^\text{anon}$ are consistently lower than the O condition decoded by $ASR_\text{eval}$. 
In contrast, for the proposed model, the original data yield the lowest WERs for most SASs, regardless of the training data of ASR evaluation models, except for the \textbf{GAN}-based SAS.
Possible reasons could be either that the $ASR_\text{eval}$ provided by VPC was not adequately trained or the structure of this model is not powerful enough.
In contrast, our proposed ASR model is more powerful in achieving accurate results.

Regarding training and evaluation time for the ASR models (Figure \ref{fig:compare-vpc-proposed-utility}), we observe that our model increases the time for ASR evaluation from 5 hours of the VPC model to 6 hours. However, the training of the proposed model takes only 20 hours, significantly less than the 72 hours required by the VPC model. Moreover, comparing the A-lazy and A-semi results show that training the ASR model on anonymized data, resulting in the $ASR_\text{eval}^\text{anon}$ model, has less effect for the proposed model than the one from the VPC. It can therefore be argued that training the $ASR_\text{eval}^\text{anon}$ model for each evaluation and using the A-semi condition is not necessary with the proposed model. This is further supported by arguing that using an ASR model specifically trained on anonymized data may be unrealistic for actual applications. Thus, by using a more robust ASR model and reverting to only the A-lazy condition, we can effectively reduce the evaluation time by 92\% from 77 hours of the VPC to 6 hours\footnote{In the A-semi condition, we have a reduction by 66\% (from 77h to 26).}.

\begin{table}[]
    \centering
    \caption{Comparisons of $\text G_\text{VD}$ obtained by the VPC and proposed ECAPA-TDNN + cosine evaluation models. The evaluation models can be either only $ASV_\text{eval}$, $ASV_\text{eval}^\text{anon}$, or a combination of both ($ASV_\text{eval}$ for original and $ASV_\text{eval}^\text{anon}$ for anonymized data).
    \texttt{\#utts per spk=5} means randomly selecting 5 utterances per speaker for similarity computation.}
    \resizebox{\linewidth}{!}{
    \begin{tabular}{ccc|rrr}
        \toprule
         SAS & Eval & \# utts per spk & $ASV_\text{eval}$ & $ASV_\text{eval}^\text{anon}$ &  Both \\
         \midrule
         \multirow{3}{*}{\textbf{BL 1.a}} & VPC & all & -7.71 & 0.18 & -1.11\\
         & Proposed & all & -6.57 & 1.37 & 0.03\\
         & Proposed & 5 & -6.79 & 1.46 & 0.11\\
         \midrule
         \multirow{3}{*}{\textbf{BL 1.b}} & VPC & all & -7.29 & 0.13 & -1.09\\
         & Proposed & all & -7.40 & 1.72 & 0.62\\
         & Proposed & 5 & -7.39 & 1.80 & 0.72\\
         \midrule
         \multirow{3}{*}{\textbf{OHNN}} & VPC & all & -1.34 & -0.11 & -1.10\\
         & Proposed & all & -1.53 & 1.69 & 0.23\\
         & Proposed & 5 & -1.57 & 1.65 & 0.15\\
         \midrule
         \multirow{3}{*}{\textbf{GAN}} & VPC & all & -0.84 & 0.68 & 0.11\\
         & Proposed & all & -1.74 & 2.64 & 2.03\\
         & Proposed & 5 & -1.62 & 2.63 & 1.96 \\
         \bottomrule
    \end{tabular}}
    \label{tab:gvd} 
\end{table}

\subsection{GAIN OF VOICE DISTINCTIVENESS}
Table \ref{tab:gvd} lists $\text G_\text{VD}$ results for various SASs computed by the VPC and proposed evaluation models.
Looking at the $\text G_\text{VD}$ achieved by $ASV_\text{eval}$, we can see:
(1) Anonymized speech generated through the \textbf{OHNN}- and \textbf{GAN}-based models exhibits higher voice distinctiveness than \textbf{BL 1.a} and \textbf{BL 1.b}, with $\text G_\text{VD}$ closer to zero.
(2) The proposed model achieves either similar or lower $\text G_\text{VD}$ values compared to the VPC model.
(3) Comparing the proposed model using all utterances to that using 5 utterances per speaker reveals similar results, suggesting that the use of 5 utterances may be sufficient for 
small test sets with limited voice variation\footnote{LibriSpeech and VCTK contain read speech segments extracted from longer recordings.}.
At the same time, using only 5 utterances per speaker reduces the time for computing the $\text G_\text{VD}$ drastically from 6 hours to only 20 minutes (Figure \ref{fig:compare-vpc-proposed-privacy}). Thus, we can speed up the $\text G_\text{VD}$ evaluation by 95\% without reducing the result quality.

However, when employing $ASV_\text{eval}^\text{anon}$ or a combination of both $ASV_\text{eval}$ and $ASV_\text{eval}^\text{anon}$, the $\text G_\text{VD}$ is significantly higher and often above zero, indicating a positive gain in voice distinctiveness. The difference between using the VPC models and the proposed ones is more notable, whereby the proposed models suggest almost no difference between the SASs anymore. This shows that $\text G_\text{VD}$ is an unstable metric that highly depends on the model used for evaluation. In order to measure voice distinctiveness precisely, it may be necessary to consider downstream tasks like 
speaker diarization \cite{anguera2012speaker}, instead of relying solely on the $\text G_\text{VD}$ metric.

\begin{figure*}
    \centering
    \begin{subfigure}[b]{0.48\linewidth}
    \begin{tikzpicture}
        \begin{groupplot}[
            group style={
                group name=group1,
                group size=3 by 1,
                xlabels at=edge bottom,
                xticklabels at=edge bottom,
                horizontal sep=0pt
            },
                ybar=3*\pgflinewidth,
                xtick = data,
                ymajorgrids=true,
                major x tick style = transparent,
                ymin=0, ymax=18,
                height=4cm,
                x=0.95cm,
                every axis y label/.style={
                at={(ticklabel cs:0.5)},rotate=90,anchor=near ticklabel,
                font=\footnotesize
                },
                tick align=outside,
                legend style={at={(2.55,1.0)},anchor=south,
                /tikz/every even column/.append style={column sep=0.2cm}, 
                font=\footnotesize},
                legend columns=2,
                clip=false
            ]

            \nextgroupplot[
            ylabel={Error Rate \%},
            symbolic x coords={all},
            ytick={0,4,8,12,16,20},
            ytick pos=left,
            bar width=10pt,
            font=\footnotesize
            ]
            \addplot[color=YellowOrange,  fill=YellowOrange!30]
            coordinates {
            (all, 7.62)
            };
            \addplot[color=RoyalBlue,  fill=RoyalBlue!30] % ASR
            coordinates {
            (all, 7.655)
            };

            \legend{ASV (EER), ASR (WER)};
            
            \nextgroupplot[
            symbolic x coords={90, 50, 10},
            ytick={0,4,8,12,16,20},
            ymajorticks=false,
            bar width=10pt,
            enlarge x limits=0.35,
            font=\footnotesize
            ]
            \addplot[color=YellowOrange,  fill=YellowOrange!30]
            coordinates {
            (90, 8.51) (50, 9.78) (10, 16.25)
            };
            \addplot[color=RoyalBlue,  fill=RoyalBlue!30] % ASR
            coordinates {
            (90, 7.775) (50, 7.8) (10, 7.91)
            };
            
            \nextgroupplot[
            symbolic x coords={500, 200},
            ytick={0,4,8,12,16,20},
            ymajorticks=false,
            bar width=10pt,
            enlarge x limits=0.65,
            font=\footnotesize
            ]
            \addplot[color=YellowOrange,  fill=YellowOrange!30]
            coordinates {
           (500, 9.23) (200, 11.50)
            }; 
            \addplot[color=RoyalBlue,  fill=RoyalBlue!30] % ASR
            coordinates {
           (500, 7.88) (200, 7.94)
            }; 
            
        \end{groupplot}
         \draw[decorate, decoration={brace, mirror, amplitude=3pt},] ([yshift=-0.8cm]2.4,0.2)-- node[below=0.1cm,font=\footnotesize]{\# utts per spk}([yshift=-0.8cm]4.6,0.2);
         \draw[decorate, decoration={brace, mirror, amplitude=3pt},] ([yshift=-0.8cm]5.5,0.2)-- node[below=0.1cm,font=\footnotesize]{\# spk}([yshift=-0.8cm]6.9,0.2);
    \end{tikzpicture}
    \subcaption{Error rates for $ASV_\text{eval}^\text{anon}$ and $ASR_\text{eval}^\text{anon}$}
    \label{fig:data_reduction_errors}
    \end{subfigure}
    \hfill
    \begin{subfigure}[b]{0.48\linewidth}
    \begin{tikzpicture}
        \begin{groupplot}[
            group style={
                group name=group1,
                group size=2 by 1,
                xlabels at=edge bottom,
                xticklabels at=edge bottom,
                horizontal sep=0pt,
                y descriptions at=edge left
            },
                xtick = data,
                ymajorgrids=true,
                major x tick style = transparent,
                ymin=0, ymax=120,
                height=4cm,
                x=0.8cm,
                every axis y label/.style={
                at={(ticklabel cs:0.5)},rotate=90,anchor=near ticklabel,font=\footnotesize
                },
                tick align=outside,
                legend pos=north west,
                legend style={fill=none},
                legend columns=2,
                clip=false
            ]
            
            \nextgroupplot[
            ylabel={Training data [\# utts]},
            symbolic x coords={all, 90, 50, 10},
            ytick={0,20, 40, 60, 80, 100},
            yticklabels={0, 20k, 40k, 60k, 80k, 100k},
            ytick pos=left,
            enlarge x limits=0.15,
            font=\footnotesize
            ]
            \addplot[ybar=3*\pgflinewidth, bar width=12pt, color=RoyalBlue,  fill=RoyalBlue!30]
            coordinates {
            (all, 104) (90, 82) (50, 46) (10, 9)
            };
            
            \nextgroupplot[
            symbolic x coords={all, 500, 200},
            ytick={0,20, 40, 60, 80, 100},
            ymajorticks=false,
            enlarge x limits=0.2,
            font=\footnotesize
            ]
            \addplot[ybar=3*\pgflinewidth, bar width=12pt, color=RoyalBlue,  fill=RoyalBlue!30]
            coordinates {
            (all, 104) (500, 57) (200, 23)
            }; 
            
        \end{groupplot}

        \begin{groupplot}[
            group style={
                group name=group1,
                group size=2 by 1,
                xlabels at=edge bottom,
                xticklabels at=edge bottom,
                horizontal sep=0pt,
                y descriptions at=edge right
            },
                xtick=\empty,
                axis line style=transparent,
                ymin=0, ymax=1200,
                height=4cm,
                x=0.8cm,
                every axis y label/.style={
                at={(ticklabel cs:0.5)},rotate=90,anchor=near ticklabel, font=\footnotesize
                },
                tick align=outside,
                clip=false,
                legend style={at={(0.9,1.0)},anchor=south,
                /tikz/every even column/.append style={column sep=0.2cm}, font=\footnotesize},
                legend columns=4
            ]
            
            \nextgroupplot[
            symbolic x coords={all, 90, 50, 10},
            ytick={0, 200, 400, 600, 800, 1000},
            yticklabels={0, 200, 400, 600, 800, 1k},
            ymajorticks=false,
            enlarge x limits=0.15,
            font=\footnotesize
            ]
            \addplot[mark=*, color=ForestGreen, mark size=1.5]
            coordinates {
            (all, 855.00) (90, 675.45) (50, 376.2) (10, 76.95)
            };
            \addplot[mark=*, color=Plum, mark size=1.5]
            coordinates {
            (all, 112.0) (90, 89.72) (50, 49.21) (10, 9.78)
            };
            \addplot[mark=*, color=Maroon, mark size=1.5]
            coordinates {
            (all, 967.0) (90, 765.17) (50, 425.41) (10, 86.73)
            };

            \legend{Anonymization, Training, Total}
            \addlegendimage{area legend, RoyalBlue, fill=RoyalBlue!30}
            \addlegendentry{Data}
            
            \nextgroupplot[
            ylabel={Time [min]},
            symbolic x coords={all, 500, 200},
            ytick={0, 200, 400, 600, 800, 1000},
            yticklabels={0, 200, 400, 600, 800, 1k},
            enlarge x limits=0.2,
            ytick pos=right,
            font=\footnotesize
            ]
            \addplot[mark=*, color=ForestGreen, mark size=1.5]
            coordinates {
            (all, 855.00) (500, 470.25) (200, 188.1)
            };
            \addplot[mark=*, color=Plum, mark size=1.5]
            coordinates {
            (all, 112.0) (500, 58.99) (200, 23.65)
            }; 
            \addplot[mark=*, color=Maroon, mark size=1.5]
            coordinates {
            (all, 967.0) (500, 529.24) (200, 211.75)
            }; 
        \end{groupplot}

        \draw[decorate, decoration={brace, mirror, amplitude=3pt},] ([yshift=-0.8cm]1.0,0.2)-- node[below=0.1cm,font=\footnotesize]{\# utts per spk}([yshift=-0.8cm]3.0,0.2);
        \draw[decorate, decoration={brace, mirror, amplitude=3pt},] ([yshift=-0.8cm]4.1,0.2)-- node[below=0.1cm,font=\footnotesize]{\# spk}([yshift=-0.8cm]5.2,0.2);
    \end{tikzpicture}
    \subcaption{Time and data efficiency for $ASV_\text{eval}^\text{anon}$}
    \label{fig:data_reduction_cost}
    \end{subfigure}
    \caption{Effect of different training strategies for $ASV_\text{eval}^\text{anon}$ and $ASR_\text{eval}^\text{anon}$ on (a) evaluation metrics for \textbf{BL 1.a}, and (b) data and time efficiency. The strategies involve finetuning with data reduction, either by restricting the number of utterances per speaker or the number of speakers. They are compared against using all data for training the models from scratch.}
       \label{fig:data_reduction}
\end{figure*}
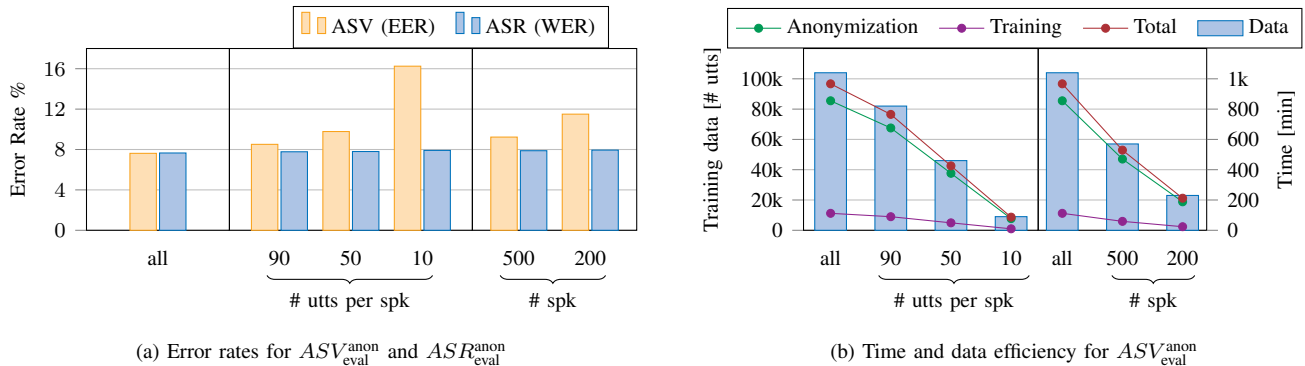

\subsection{TRAINING STRATEGY FOR $ASV_\text{eval}^\text{anon}$ AND $ASR_\text{eval}^\text{anon}$ MODELS}
Figure \ref{fig:data_reduction} shows the influence of the different data reduction strategies on the privacy scores and evaluation efficiency for \textbf{BL 1.a}. 
The experiments are conducted across various $ASV_\text{eval}^\text{anon}$, trained using different amounts of anonymized speech data.

It can be observed from the results in Figure \ref{fig:data_reduction_errors} that the more we decrease the amount of training data, the more the EER increases. This is especially problematic for \texttt{\#utts per spk=10} because its scores might suggest that the SAS's privacy protection would be better than it actually is. 
For WER, the effect of data reduction is negligible as it only changes from 7.66\% WER (\texttt{all}) to 7.91\% (\texttt{\#utts per spk=10}) in the worst case.

The biggest impact of the different training strategies for $ASV_\text{eval}^\text{anon}$ can be seen in Figure \ref{fig:data_reduction_cost}. It shows that the largest factor of time needed for privacy evaluation comes from the requirement of having to anonymize the training data for each evaluation run. Decreasing the amount of training data therefore means less time being spent on anonymizing it, thus, the time cost of an evaluation decreases linearly with the reduction of data. 

Based on these results, we found that \texttt{\#utts per spk=50} provides the best balance between EER increase and cost\footnote{We validated this finding using other SASs as well. We omitted them due to limited space.}. This data reduction decreases the total evaluation time for ASV evaluation from 16 hours (14.25 hours for anonymization, 2 hours for training) to 7 hours (6.25 hours for anonymization, 49 minutes for training). Compared to 20.25 hours needed in the VPC evaluation framework (14.25 hours for anonymization, 6 hours for training), this reduces the time needed for creating the $ASV_\text{eval}^\text{anon}$ model by 65\%. Therefore, we recommend to use finetuning with this setting at least during SAS development, and only revert to training on all data from scratch for final evaluation.

\section{ANALYSIS} 
This section delves deeper into privacy results by examining resynthesis performance and summarizing rankings when employing various evaluation models for different SASs.

\begin{table}[]
    \centering
    \caption{Comparison of EERs (\%) and WERs (\%) on the resynthesized conditions without anonymization. The proposed models are ECAPA-TDNN + cosine for ASV and transformer-based CTC/AED ASR.}
    \resizebox{\linewidth}{!}{
    \begin{tabular}{cc|rrr|rr}
        \toprule
        & & \multicolumn{3}{c|}{EER (\%)} & \multicolumn{2}{c}{WER (\%)}\\
        SAS & Eval & OO & OR & RR-lazy & O & R-lazy\\
        \midrule
        \multirow{2}{*}{\textbf{BL 1.a}} & VPC & 3.58 & 17.53 & 13.24 & 8.48 & 10.06\\ 
        & Proposed & 3.11 & 22.47 & 14.10 & 7.24 & 11.00\\ 
        \midrule
        \multirow{2}{*}{\textbf{BL 1.b}} & VPC & 3.58 & 21.25 & 14.62 & 8.48 & 12.73\\ 
        & Proposed & 3.11 & 28.45 & 15.28 & 7.24 & 9.75\\ 
        \midrule
        \multirow{2}{*}{\textbf{OHNN}} & VPC & 3.30 & 7.60 & 4.82 & 8.48 & 9.45\\ 
        & Proposed & 3.11 & 10.05 & 4.28 & 7.24 & 7.58\\ 
        \midrule
        \multirow{2}{*}{\textbf{GAN}} & VPC & 3.29 & 23.09 & 16.08 & 8.48 & 8.16\\ 
        & Proposed  & 3.11 & 27.09 & 17.68 & 7.24 & 7.85\\ 
        \bottomrule
    \end{tabular}}
    \label{tab:resynthesis_all}
\end{table}

\subsection{RESYNTHESIS}
In the privacy evaluation, we obtain one privacy score (EER) for each model and attack condition. However, as all tested SASs employ a speech synthesis step after the actual anonymization method, it is not clear whether the anonymization power of an SAS actually from this anonymization method or from the synthesis. We therefore explore testing the \textit{resynthesis performance} of each SAS. For this, we generate a new version of the evaluation data per SAS in which the anonymization method is skipped and instead the original speaker vector is used for synthesis. We evaluate two new conditions for $ASV_\text{eval}$: OR and RR-lazy, with original or resynthesized enrollment data, respectively, and resynthesized trial data. For $ASR_\text{eval}$, we test the decoding performance on the resynthesized data (R-lazy). In both cases, we compare to the performance on the original data. 

Table \ref{tab:resynthesis_all} shows the results. Except for the \textbf{OHNN}-based SAS, all SASs clearly exploit the synthesis to increase the privacy protection and do not rely only on their anonymization method. The \textbf{OHNN}-based SAS, on the other hand, has almost no identification loss during resynthesis. However, the synthesis in all SASs lead to an increase in WER and thus reduced intelligibility. 

Using the synthesis to increase the privacy protection is not necessarily a drawback of the \textbf{BL 1.a}, \textbf{BL 1.b} and \textbf{GAN}-based SAS. However, researchers might not be aware of this effect and might put too much focus on optimizing their anonymization method instead of the synthesis. It also decreases the control one has about the actual outcome of the SAS. A related issue was already observed by \cite{panariello23vocoder} about the vocoder drift of speaker vectors during anonymization.

\subsection{EFFECT ON RANKING}
In this paper, we presented new evaluation models and strategies for training the \textit{lazy-informed} attackers as alternatives to the evaluation framework of the VPC. We have shown that the perceived strength of an SAS depends on the models it has been evaluated with, however, it is also important to analyze how the relative performance of multiple SAS in comparison (i.e., their ranking) is influenced by the choice of evaluation models.

Comparing the scores for different SAS in the \textit{lazy-} and \textit{semi-informed} attack conditions as shown in Table \ref{tab:res_eval_models_summary} reveals consistently the same ranking for the level of privacy protection (with \textbf{GAN}- and \textbf{OHNN}-based being partly on equal places): (1) \textbf{GAN}-based, (2) \textbf{OHNN}-based, (3) \textbf{BL 1.a}, and (4) \textbf{BL 1.b}. This is regardless of whether the VPC evaluation models or the proposed ones are applied, and also regardless of the training strategy. For ASR evaluation, on the other hand, the ranking of SASs' performances does not stay consistent but changes depending on the ASR model used for evaluation. However, the WER scores of all SASs are relatively similar to each other when using the proposed evaluation models. Thus, we argue that this change in ranking for utility evaluation is a rather small effect.

\section{DISCUSSION}

\textit{Which evaluation models should we consider?}
For privacy evaluation, we proposed and evaluated various attacker models against a selection of SAS and found that the choice of attack model did not influence the ranking of privacy protection ability for the chosen SASs, although they produced different privacy scores.
However, we only tested a limited number of attackers from the same ASV family.
It is possible that an attacker using a different technique, e.g., conformer-based \cite{zhang22h_interspeech} or SSL-based ASV models \cite{chen2022unispeech}, might result in a different trend.

Moreover, the choice of attack conditions is still heavily based on assumptions. 
It is unclear whether \textit{semi-informed} attackers are realistic or what we could assume about the knowledge and dedication of real attackers. Hopefully, challenges like a Voice Privacy Attacker Challenge\footnote{A Voice Privacy Attacker Challenge was initially planned for a workshop at INTERSPEECH 2023, see \url{https://www.voiceprivacychallenge.org}.} will provide new insights and perspectives.

Overall, we saw a particular trade-off between the quality of objective results and their usability in terms of time requirement for computation, at least for privacy metrics. Retraining the $ASV_\text{eval}^\text{anon}$ from scratch on the full anonymized training data seems to lead to the strongest attacker. However, it is costly, which can be significantly reduced by a finetuning strategy that leads to  
minimal reduction in attacker performance. 
We propose using this alternative training strategy during SAS development to speed up voice privacy research. However, a full retraining on all data might still be a better option for a final assessment of the full privacy capabilities.

\textit{What are the drawbacks of the current evaluation metrics?}
According to our experiments, $\text G_\text{VD}$ is a more problematic metric. We tested three model approaches ($ASV_\text{eval}$, $ASV_\text{eval}^\text{anon}$, and the combination of both), but their results differ considerably.
It is unclear which approach is better suited for measuring the preservation of voice distinctiveness among anonymized speech.
We conclude that we need a more robust alternative, e.g., to use the anonymized dataset to perform downstream speaker verification or speaker diarization tasks.

\textit{What is missing?}
Currently, there are no definitions or measurable criteria for success or the guarantee of full privacy protection through anonymization since all existing evaluations rely on assumptions and specific attack models.
It is unknown when an SAS could be considered good enough for use on data where privacy protection matters, or how the remaining privacy risk of current systems can be accurately measured\footnote{A first approach for privacy risk assessment was proposed in the ZEBRA framework \cite{nautsch2020zebra} but is so far not used as a standard metric in research.}.

\textit{In summary}, being open source, the proposed framework serves as a platform for unifying researchers and research on this topic. Researchers can add their own SASs and evaluation metrics to the framework such that large-scale and extensive evaluations and comparisons would be possible without much additional effort. In this way, we hope that this framework will help towards finding answers to the questions above, and towards the development of powerful anonymization tools.

\section{CONCLUSION} 
We proposed a new Python-based and modular open-source framework for speaker anonymization research. It allows combining, managing, and evaluating several anonymization approaches within one platform that is simple to apply and extend. We further present various improvements to standard evaluation techniques for speaker anonymization. Specifically, we exchange previous Kaldi-based evaluation models with more powerful techniques using the ESPnet and SpeechBrain toolkits. Moreover, we showed that we could decrease the time required for evaluation by up to 95\% by reducing training and test data while keeping the quality of the evaluations at compatible levels. We anticipate that these changes to common development and evaluation procedures will significantly facilitate and support speaker anonymization research in the near future.

\bibliographystyle{IEEEbib}
\bibliography{mybib}

\begin{thebibliography}{10}

\bibitem{tomashenko2021voiceprivacy}
N.~Tomashenko, X.~Wang, E.~Vincent, J.~Patino, B.~M.~L. Srivastava, P.-G.
  No{\'e}, A.~Nautsch, N.~Evans, J.~Yamagishi, B.~O'Brien, et~al.,
\newblock ``The {VoicePrivacy} 2020 challenge: Results and findings,''
\newblock {\em Computer Speech \& Language}, vol. 74, pp. 101362, Jul 2022.

\bibitem{povey2018semi}
D.~Povey, G.~Cheng, Y.~Wang, K.~Li, H.~Xu, M.~Yarmohammadi, and S.~Khudanpur,
\newblock ``Semi-orthogonal low-rank matrix factorization for deep neural
  networks.,''
\newblock in {\em Proc. Interspeech}, 2018, pp. 3743--3747.

\bibitem{tomashenko2022voiceprivacy}
N.~Tomashenko, X.~Wang, X.~Miao, H.~Nourtel, P.~Champion, M.~Todisco,
  E.~Vincent, N.~Evans, J.~Yamagishi, and J.~F. Bonastre,
\newblock ``The {VoicePrivacy} 2022 challenge evaluation plan,''
\newblock {\em arXiv preprint arXiv:2203.12468}, 2022.

\bibitem{kaldi}
D.~Povey, A.~Ghoshal, G.~Boulianne, L.~Burget, O.~Glembek, N.~Goel,
  M.~Hannemann, P.~Motlicek, Y.~Qian, P.~Schwarz, J.~Silovsky, G.~Stemmer, and
  K.~Vesely,
\newblock ``The kaldi speech recognition toolkit,''
\newblock in {\em Proc. IEEE ASRU}, Dec. 2011.

\bibitem{franzreb2023comprehensive}
C.~Franzreb, T.~Polzehl, and S.~Moeller,
\newblock ``{A Comprehensive Evaluation Framework for Speaker Anonymization
  Systems},''
\newblock in {\em Proc. 3rd Symposium on Security and Privacy in Speech
  Communication}, 2023, pp. 65--72.

\bibitem{zhang2023voicepm}
Z.~Shaohu, L.~Zhouyu, and D.~Anupam,
\newblock ``Voicepm: A robust privacy measurement on voice anonymity,''
\newblock in {\em Proc. 16th ACM Conference on Security and Privacy in Wireless
  and Mobile Networks (WiSec)}, 2023, p. 215–226.

\bibitem{espnet}
S.~Watanabe, T.~Hori, S.~Karita, T.~Hayashi, J.~Nishitoba, Y.~Unno, N.~{Enrique
  Yalta Soplin}, J.~Heymann, M.~Wiesner, N.~Chen, A.~Renduchintala, and
  T.~Ochiai,
\newblock ``{ESPnet}: End-to-end speech processing toolkit,''
\newblock in {\em Proc. Interspeech}, 2018, pp. 2207--2211.

\bibitem{speechbrain}
M.~Ravanelli, T.~Parcollet, P.~Plantinga, A.~Rouhe, S.~Cornell, L.~Lugosch,
  C.~Subakan, N.~Dawalatabad, A.~Heba, J.~Zhong, J.-C. Chou, S.-L. Yeh, S.-W.
  Fu, C.-F. Liao, E.~Rastorgueva, F.~Grondin, W.~Aris, H.~Na, Y.~Gao, R.~D.
  Mori, and Y.~Bengio,
\newblock ``{SpeechBrain}: A general-purpose speech toolkit,''
\newblock {\em arXiv preprint arXiv:2106.04624}, 2021.

\bibitem{meyer22b_interspeech}
S.~Meyer, F.~Lux, P.~Denisov, J.~Koch, P.~Tilli, and N.~T. Vu,
\newblock ``Speaker anonymization with phonetic intermediate representations,''
\newblock in {\em Proc. Interspeech}, 2022, pp. 4925--4929.

\bibitem{meyer2023anonymizing}
S.~Meyer, P.~Tilli, P.~Denisov, F.~Lux, J.~Koch, and N.~T. Vu,
\newblock ``Anonymizing speech with generative adversarial networks to preserve
  speaker privacy,''
\newblock in {\em Proc. IEEE SLT}. IEEE, 2023, pp. 912--919.

\bibitem{meyer2023prosody}
S.~Meyer, F.~Lux, J.~Koch, P.~Denisov, P.~Tilli, and N.~T. Vu,
\newblock ``Prosody is not identity: A speaker anonymization approach using
  prosody cloning,''
\newblock in {\em Proc. IEEE ICASSP}. IEEE, 2023, pp. 1--5.

\bibitem{miao22_odyssey}
X.~Miao, X.~Wang, E.~Cooper, J.~Yamagishi, and N.~Tomashenko,
\newblock ``Language-independent speaker anonymization approach using
  self-supervised pre-trained models,''
\newblock in {\em Proc. Odyssey}, 2022, pp. 279--286.

\bibitem{miao2022analyzing}
X.~Miao, X.~Wang, E.~Cooper, J.~Yamagishi, and N.~Tomashenko,
\newblock ``Analyzing language-independent speaker anonymization framework
  under unseen conditions,''
\newblock in {\em Proc. Interspeech}, 2022, pp. 4426--4430.

\bibitem{miao2023language}
X.~Miao, X.~Wang, E.~Cooper, J.~Yamagishi, and N.~Tomashenko,
\newblock ``Speaker anonymization using orthogonal householder neural
  network,''
\newblock {\em IEEE/ACM Trans. Audio, Speech, and Language Processing}, vol.
  31, pp. 3681--3695, 2023.

\bibitem{mawalim2022speaker}
C.~O. Mawalim, K.~Galajit, J.~Karnjana, S.~Kidani, and M.~Unoki,
\newblock ``Speaker anonymization by modifying fundamental frequency and
  x-vector singular value,''
\newblock {\em Computer Speech \& Language}, vol. 73, pp. 101326, 2022.

\bibitem{champion2022disentangled}
P.~Champion, A.~Larcher, and D.~Jouvet,
\newblock ``Are disentangled representations all you need to build speaker
  anonymization systems?,''
\newblock in {\em Proc. Interspeech}, 2022, pp. 2793--2797.

\bibitem{shamsabadi2022differentially}
A.~S. Shamsabadi, B.~M.~L. Srivastava, A.~Bellet, N.~Vauquier, E.~Vincent,
  M.~Maouche, M.~Tommasi, and N.~Papernot,
\newblock ``Differentially private speaker anonymization,''
\newblock {\em {Proc. Privacy Enhancing Technologies}}, vol. 2023, no. 1, pp.
  98--114, Jan. 2023.

\bibitem{turner2022generating}
H.~Turner, G.~Lovisotto, and I.~Martinovic,
\newblock ``Generating identities with mixture models for speaker
  anonymization,''
\newblock {\em Computer Speech \& Language}, vol. 72, pp. 101318, 2022.

\bibitem{yao22_spsc}
J.~Yao, Q.~Wang, L.~Zhang, P.~Guo, Y.~Liang, and L.~Xie,
\newblock ``{NWPU-ASLP} system for the voiceprivacy 2022 challenge,''
\newblock in {\em Proc. 2nd Symp. on Security and Privacy in Speech
  Communication}, 2022.

\bibitem{peddinti2015time}
V.~Peddinti, D.~Povey, and S.~Khudanpur,
\newblock ``A time delay neural network architecture for efficient modeling of
  long temporal contexts,''
\newblock in {\em Proc. Interspeech}, 2015, pp. 3214--3218.

\bibitem{kasi2002yet}
K.~Kasi and S.~A. Zahorian,
\newblock ``Yet another algorithm for pitch tracking,''
\newblock in {\em Proc. IEEE ICASSP}, 2002, vol.~1, pp. 361--364.

\bibitem{snyder2018x}
D.~Snyder, D.~Garcia-Romero, G.~Sell, D.~Povey, and S.~Khudanpur,
\newblock ``X-vectors: Robust dnn embeddings for speaker recognition,''
\newblock in {\em Proc. IEEE ICASSP}. IEEE, 2018, pp. 5329--5333.

\bibitem{desplanques2020ecapa}
B.~Desplanques, J.~Thienpondt, and K.~Demuynck,
\newblock ``{ECAPA-TDNN}: Emphasized channel attention, propagation and
  aggregation in {TDNN} based speaker verification,''
\newblock in {\em Proc. Interspeech}, 2020, pp. 3830--3834.

\bibitem{Srivastava2020DesignCF}
B.~M.~L. Srivastava, N.~Tomashenko, X.~Wang, E.~Vincent, J.~Yamagishi,
  M.~Maouche, A.~Bellet, and M.~Tommasi,
\newblock ``Design choices for x-vector based speaker anonymization,''
\newblock in {\em Proc. Interspeech}, 2020, pp. 1713--1717.

\bibitem{wang2019neural}
X.~Wang, S.~Takaki, and J.~Yamagishi,
\newblock ``Neural source-filter-based waveform model for statistical
  parametric speech synthesis,''
\newblock in {\em Proc. IEEE ICASSP}, 2019, pp. 5916--5920.

\bibitem{kong2020hifi}
J.~Kong, J.~Kim, and J.~Bae,
\newblock ``{HiFi-GAN}: Generative adversarial networks for efficient and high
  fidelity speech synthesis,''
\newblock in {\em Proc. NeurIPS}, 2020, pp. 17022--17033.

\bibitem{gomez-barrero2018general}
M.~Gomez-Barrero, J.~Galbally, C.~Rathgeb, and C.~Busch,
\newblock ``General framework to evaluate unlinkability in biometric template
  protection systems,''
\newblock {\em IEEE Trans. Information Forensics and Security}, vol. 13, no. 6,
  pp. 1406--1420, 2018.

\bibitem{maouche2020a-comparative}
M.~Maouche, B.~M.~L. Srivastava, N.~Vauquier, A.~Bellet, M.~Tommasi, and
  E.~Vincent,
\newblock ``A comparative study of speech anonymization metrics,''
\newblock in {\em Proc. Interspeech}, 2020, pp. 1708--1712.

\bibitem{noe2020speech}
P.-G. Noé, J.-F. Bonastre, D.~Matrouf, N.~Tomashenko, A.~Nautsch, and
  N.~Evans,
\newblock ``Speech pseudonymisation assessment using voice similarity
  matrices,''
\newblock in {\em Proc. Interspeech}, 2020, pp. 1718--1722.

\bibitem{nautsch2020zebra}
A.~Nautsch, J.~Patino, N.~Tomashenko, J.~Yamagishi, P.-G. Noé, J.-F. Bonastre,
  M.~Todisco, and N.~Evans,
\newblock ``The privacy {ZEBRA}: Zero evidence biometric recognition
  assessment,''
\newblock in {\em Proc. Interspeech}, 2020, pp. 1698--1702.

\bibitem{pytorch}
A.~Paszke, S.~Gross, F.~Massa, A.~Lerer, J.~Bradbury, G.~Chanan, T.~Killeen,
  Z.~Lin, N.~Gimelshein, L.~Antiga, A.~Desmaison, A.~Kopf, E.~Yang, Z.~DeVito,
  M.~Raison, A.~Tejani, S.~Chilamkurthy, B.~Steiner, L.~Fang, J.~Bai, and
  S.~Chintala,
\newblock ``{PyTorch}: An imperative style, high-performance deep learning
  library,''
\newblock in {\em Advances in Neural Information Processing Systems 32}, pp.
  8024--8035. Curran Associates, Inc., 2019.

\bibitem{watanabe2017hybrid}
S.~Watanabe, T.~Hori, S.~Kim, J.~R. Hershey, and T.~Hayashi,
\newblock ``Hybrid {CTC}/attention architecture for end-to-end speech
  recognition,''
\newblock {\em IEEE Journal of Selected Topics in Signal Processing}, vol. 11,
  no. 8, pp. 1240--1253, 2017.

\bibitem{panayotov2015librispeech}
V.~Panayotov, G.~Chen, D.~Povey, and S.~Khudanpur,
\newblock ``Librispeech: An {ASR} corpus based on public domain audio books,''
\newblock in {\em Proc. IEEE ICASSP}, 2015, pp. 5206--5210.

\bibitem{yamagishi2019vctk}
J.~Yamagishi, C.~Veaux, and K.~MacDonald,
\newblock ``{CSTR VCTK} corpus: English multi-speaker corpus for {CSTR} voice
  cloning toolkit (version 0.92),'' 2019.

\bibitem{wang22r_interspeech}
Q.~Wang, K.~A. Lee, and T.~Liu,
\newblock ``Scoring of large-margin embeddings for speaker verification: Cosine
  or {PLDA}?,''
\newblock in {\em Proc. Interspeech}, 2022, pp. 600--604.

\bibitem{li2023make}
Z.~Li, R.~Xiao, H.~Chen, Z.~Zhao, W.~Wang, and P.~Zhang,
\newblock ``How to make embeddings suitable for plda,''
\newblock {\em Computer Speech \& Language}, vol. 81, pp. 101523, 2023.

\bibitem{anguera2012speaker}
X.~Anguera, S.~Bozonnet, N.~Evans, C.~Fredouille, G.~Friedland, and O.~Vinyals,
\newblock ``Speaker diarization: A review of recent research,''
\newblock {\em IEEE/ACM Trans. Audio, Speech, and Language Processing}, vol.
  20, no. 2, pp. 356--370, 2012.

\bibitem{panariello23vocoder}
M.~Panariello, M.~Todisco, and N.~Evans,
\newblock ``Vocoder drift in x-vector–based speaker anonymization,''
\newblock in {\em Proc. Interspeech}, 2023, pp. 2863--2867.

\bibitem{zhang22h_interspeech}
Y.~Zhang, Z.~Lv, H.~Wu, S.~Zhang, P.~Hu, Z.~Wu, H.-y. Lee, and H.~Meng,
\newblock ``{MFA-Conformer}: Multi-scale feature aggregation conformer for
  automatic speaker verification,''
\newblock in {\em Proc. Interspeech}, 2022, pp. 306--310.

\bibitem{chen2022unispeech}
S.~Chen, Y.~Wu, C.~Wang, Z.~Chen, Z.~Chen, S.~Liu, J.~Wu, Y.~Qiao, Furu W.,
  J.~Li, and X.~Yu,
\newblock ``Unispeech-sat: Universal speech representation learning with
  speaker aware pre-training,''
\newblock in {\em Proc. IEEE ICASSP}. IEEE, 2022, pp. 6152--6156.

\end{thebibliography}

\vfill\pagebreak

\end{document}